 \documentclass[twocolumn,prb,aps,showpacs]{revtex4}

 \usepackage[dvips]{graphicx}
 \usepackage[dvips]{graphics}
\usepackage{isolatin1}

\begin{document}

\title{The metal-insulator transition of the ferromagnetic La$_{7/8}$Sr$_{1/8}$MnO$_3$ probed by spin waves: a 2D stripe superstructure}
\author{M. Hennion$^1$, F. Moussa$^1$, P. Lehouelleur$^1$, P. Reutler$^2$, A. Revcolevschi$^2$}

\affiliation{$^1$ Laboratoire L\'eon Brillouin, CEA-CNRS, CE-Saclay, F-91191 Gif sur Yvette Cedex, France.\\
$^2$ Laboratoire de Physico-Chimie des Solides, Universit\'e Paris-Sud, F-91405 Orsay Cedex, France.}

\begin{abstract}
We report a study of spin-waves in hole-doped ferromagnetic La$_{7/8}$Sr$_{1/8}$MnO$_3$,  in the metallic state (165K) below T$_C$ (181K), and through the puzzling metal-insulator transition which occurs at T$_{O'O"}$=159K. They reveal very unusual excitations. Propagating spin waves are observed in the small q-range up to q=0.25 ($\lambda$=4a) and, beyond, four dispersionless levels. Both types of excitations have a quasi two-dimensional (2D) character. The transition is revealed by a folding of the dispersed magnon branch at q=1/8. In the metallic state, the dispersionless levels reveal ferromagnetic domains with 4 lattice spacings for the size along {\bf a} and {\bf b}. They lead to a picture of charge segregation with hole-poor domains surrounded with hole-rich paths. Within this description, the transition appears as the ordering of domains, which can be interpreted in terms of a 2D supertructure of orthogonal stripes.   
\end{abstract}
\maketitle

In manganites, the enhancement of ferromagnetism by decreasing temperature or by increasing the hole-doping rate, is generally associated with an increase of the metallicity. Therefore the metal-insulator transition which occurs at T$_{O'O"}$ (159K) below T$_C$ (T$_C$=181K) in La$_{7/8}$Sr$_{1/8}$MnO$_3$ remains a very puzzling problem. At T$_{O'O"}$, structural changes also occur\cite{Pinsard,Cox} giving rise to new superstructures\cite{Yamada}. Various models have been proposed to describe the low-temperature state, such as charge ordering with polaron ordering\cite{Yamada}, or a new orbital ordering\cite{Endoh,Geck}. On the theoretical side, a picture has been proposed where planes with hole-rich stripes distant from 2a, alternate along {\bf c} with planes without holes\cite{Mizokawa}. 
But the very nature of the low-temperature transition is still unsolved. A first study of La$_{7/8}$Sr$_{1/8}$MnO$_3$ by inelastic neutron scattering\cite{Moussa} mainly in the low temperature insulating state, has revealed an unusual split spin wave dispersion branch, with dispersed magnons branches lying on phonon branches, suggesting a peculiar magnon-phonon coupling. We report here inelastic neutron scattering measurements in the {\it metallic} state of this compound and their evolution {\it through the metal-insulator transition}. Below T$_C$, in the metallic phase, unusual magnetic excitations are observed, interplay of dispersed spin waves and q-independent magnetic excitations with a quasi two-dimensional character (2D). Several new striking features are observed with respect to La$_{1-x}$Ca$_{x}$MnO$_3$\cite{Hennion}, so that a simple analysis can be proposed. A picture of two-dimensional (2D) ferromagnetic domains induced by charge segregation with a limited size of 4 lattice parameters (4a) along {\bf a} and {\bf b}  crystallographic axes is suggested. The puzzling metal-insulator transition taking place at T$_{O'O"}$, below T$_C$, is seen as the appearance of a new magnetic periodicity of 4 lattice spacings along {\bf a} and {\bf b} axes for long-range spin-waves. This leads to a description of the magnetic state in terms of 2D stripes supertructure below the T$_{O'O"}$ transition.\\ 

Inelastic neutron scattering measurements have been performed on triple-axis spectrometers at the Orph\'ee reactor of the Laboratoire L\'eon Brillouin. We use a pseudo-cubic indexing of the perovskite structure to label wave-vectors, so that [100], [010] and [001] directions of the reciprocal space correspond to {\bf a}, {\bf b}, {\bf c} axis respectively.
 {\bf q} (with $\zeta$ components in the legend of figures) is defined in reduced lattice units (rlu), or in $\AA^{-1}$ when comparing several directions
in the same figure. Our sample is twinned, so that directions of distinct domains are superimposed. Fortunately, previous studies prove the equivalence of {\bf a} and {\bf b} for the magnetic coupling\cite{Moussa0} yielding [100]=[010]$\not\equiv$[001] and [101]=[011]$\not\equiv$[110]. Inelastic scans have been performed at constant q and constant outcoming neutron wavevector $k_f$ and fitted to lorentzian line-shapes convoluted with the spectrometer resolution function. E(q), $\Gamma$(q) and I(q) correspond to the energy, damping and integrated intensity of a magnetic mode. In the fits of magnetic energy spectra with several peaks, the damping value is maintained equal for all peaks. In addition, transverse (TA), longitudinal (LA) acoustic phonon modes and lower optical one (LO) have been also measured. 
Magnetic modes have been measured either around $\tau$=(001) for low energy transfers or around $\tau$=(110) for large ones. For the whole magnetic spectrum measured along [001], around $\tau$=(001), the phonon contamination is negligible, as also found in La$_{0.9}$Sr$_{0.1}$MnO$_3$\cite{Moussa}. This has been further checked by polarised neutron experiments. In contrast, phonon contamination is significant in measurements around $\tau$=(110) and also around $\tau$=(110) at large q along [111]. As shown below, in Fig. 5, in all cases, the ambiguity on the nature of the mode can be lift by studying the temperature dependence, either at room temperature (smearing of the magnetic peaks) or at low temperature (increase of intensity and dispersion). Due to a possible irreversibility, the measurements were performed in decreasing temperature.
\begin{figure}[t]
\centerline{\includegraphics[width=7 cm]{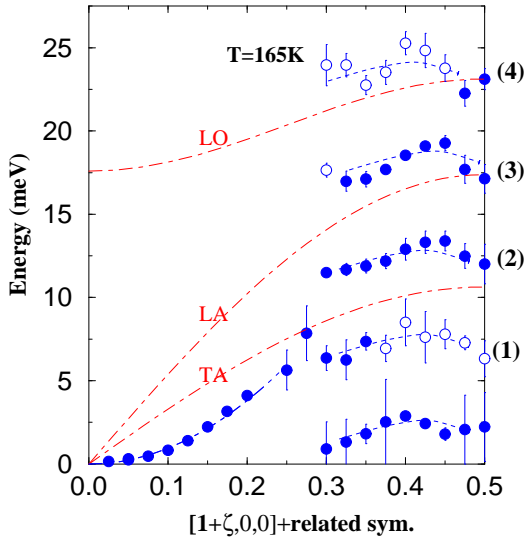}}
\caption{Spin wave spectrum along [100]+[010]+[001] at T=165K showing 4 energy levels. Full (empty) circles indicate modes with main (weak) intensity. Dot-dashed lines are fits of TA, LA and LO phonon dispersions measured at 165K around $\tau$=(200) and dashed lines are guides for the eyes.}
\end{figure}
 \begin{figure}[t]
\centerline{\includegraphics[width=7.7 cm]{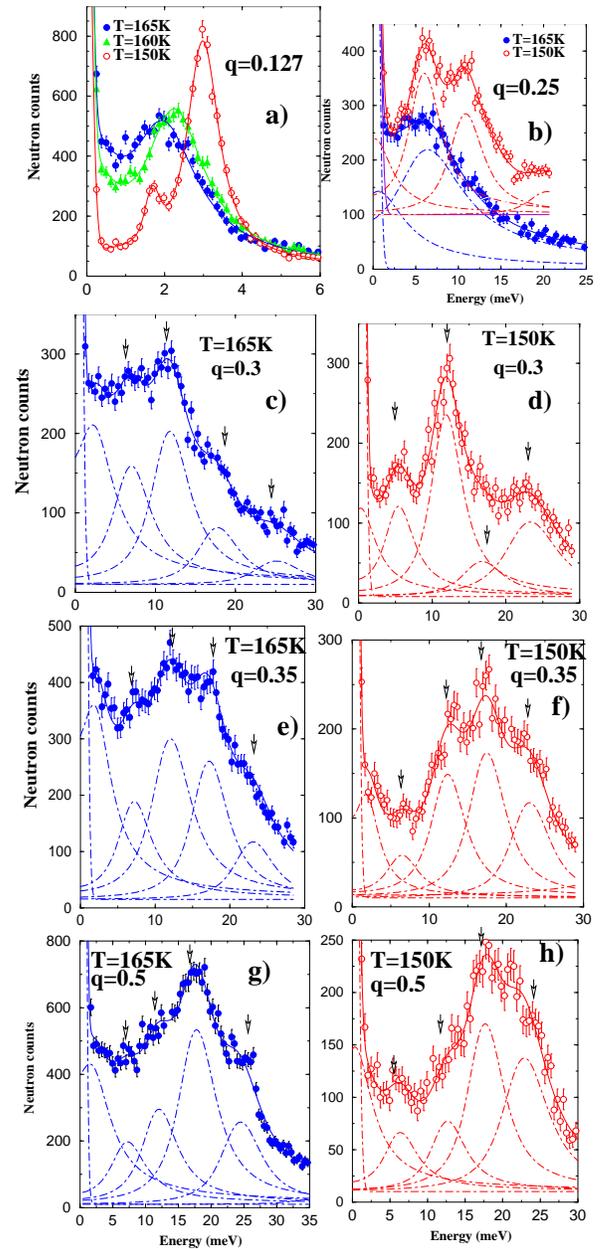}}
\caption{Examples of fitted energy spectra for q along [100]:- a) in the low-q range at q=0.127, T=165K, 160K and 150K - b) to h) in the large q range, at T=165K for q=0.25, 0.30, 0.35, 0.5 (blue filled circles in b), c), e), g)), and at T=150K for the same q values (red empty circles in b), d), f), h)). In b), where two spectra are superimposed the spectrum at 150K has been shifted by a constant intensity. Arrows indicate the energies corresponding to various levels.}
\end{figure}

{\bf Spin waves in the metallic state (T=165K)}\\
The spin wave spectrum obtained at T=165K along [100] is reported in Fig. 1. It clearly evidences the unusual features mentioned above: a dispersed spin wave branch, up to the characteristic value q=1/4 (rlu), or a wavelength $\lambda$=4a, and, beyond, nearly dispersionless levels spreading in a higher energy-range. A quasi-elastic signal ($\Delta$E$\approx$2meV) is also present. 
The separation between the (q,E) ranges of the dispersed and undispersed excitations indicates some interaction between these two types of excitations.  Examples of energy spectra at T=165K for q=0.127 and 0.25 are shown in Fig. 2-a)-b) (full circles). In the large q range, the striking feature is the existence of several levels at 7, 12.5, 18, and 24 meV, labelled (1) to (4) in Fig. 1. The main intensity is distributed on the three lower ones as q increases, the higher level being weak. Corresponding data at q=0.3, 0.35 and 0.5 are shown in Fig. 2 c), e), g). Actually, the energy spectra have been measured several times to check their reproducibility. Some differences may appear in the intensity of the various energy components specially for the quasielastic peak, but not in the value of the energy. In our previous partial temperature study\cite{Moussa}, at q=0.5, due to the truncated low energy part of the scan or the insufficient statistics, the lower levels could not be detected. In fact measurements above T$_C$ have already pointed out their existence, while the dispersed branch appears at and below T$_C$. This latter presents a significant damping ($\Gamma$(q)/E(q)$\approx$1, see  Fig. 2-a), indicative of strong disorder, while the levels have only a moderate damping with a $\Gamma$(q)/E(q) ratio spreading from $\approx$1/3 to 1/7 depending on energy. Application of a field H=3.5 T at q=0.25 (Fig. 3, left panel), induces an effect similar to the evolution observed with decreasing temperature (cf below). The significant increase of the energy, far from the expected Zeeman effect, indicates a high sensitivity of the coupling to spin disorder. It suggests that the coupling involved in the dispersed branch could be related to double exchange. We outline that in the polarised neutron experiment with polarisation analysis of Fig. 3, the data under applied field correspond to the spin-flip chanel which characterizes transverse magnetic excitations. The non-spin-flip chanel has no measurable intensity, confirming the absence of phonon contribution.
 \begin{figure}[t]
\centerline{\includegraphics[width=9 cm]{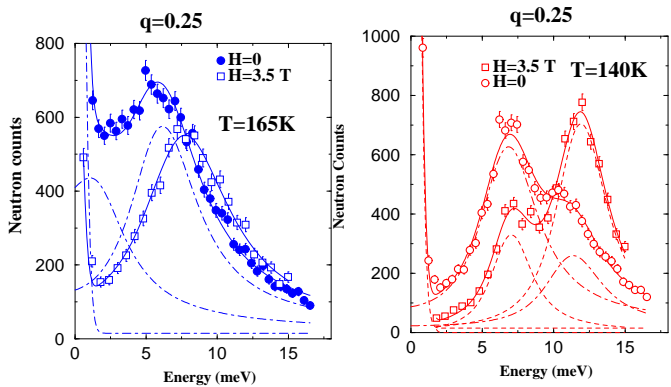}}
\caption{Field effect on the mode q=0.25, along [100]. Left panel: energy spectra at T=165K, in zero field (filled circle) and under 3.5 T (empty square). Right panel: energy spectra at T=140K, in zero field (empty circle) and under 3.5 T (empty square). Experiments under applied field correspond to the spin-flip chanel in a polarised neutron experiment with polarisation analysis, performed at the Institut Laue-Langevin on the spectrometer IN22.}
\end{figure}  
\begin{figure}[t]
\centerline{\includegraphics[width=7 cm]{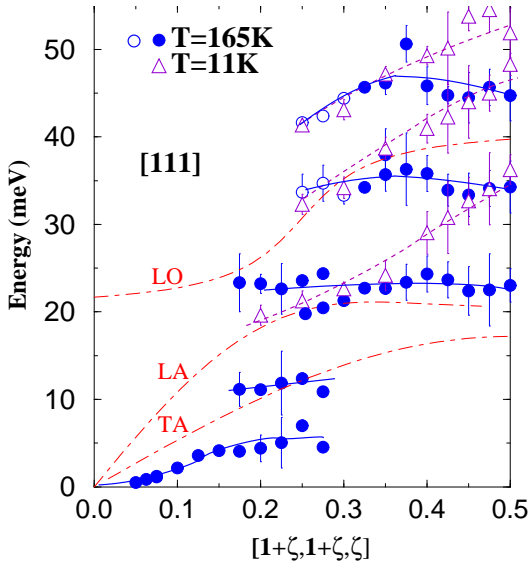}}
\caption{Spin wave spectrum along [111] at T=165K (full circles) and 11K (empty triangles) showing several levels. The temperature evolution of the 3 upper ones only is shown for clarity, confirming their magnetic origin. Full (empty) circles indicate modes with main (weak) intensity. Dot-dashed lines are fits of TA, LA and LO phonon branches measured at 250K. Solid and dashed lines are guides for the eyes.}
\end{figure}

In other directions, similar spin wave spectra with propagating and dispersion-less magnetic modes are also observed. Along [110]+[011] (2 distinct domains), not shown, the same levels and levels at twice these values are observed, except the 12.5 meV one, which is shifted at a higher energy, lying on a part of an acoustic phonon branch. Along [111] for which all domains are equivalent, at least 4 levels are determined (cf Fig. 4) above the dispersed branch. {\it Very interestingly, a factor 2, exists between their energy values and the ones determined along [100]}. Examples of spectra at q=0.25 measured around $\tau$=(001) for low energy transfers and at q=0.325 measured around $\tau$=(110) for high energy transfers, are displayed in Fig. 5-a) and 5-c) respectively. The temperature evolution (cf captions of Fig. 4, Fig. 5-c) and Fig. 5-d)) allows to ascertain their magnetic nature.\\ 
These dispersionless levels indicate local modes, interpreted below as standing spin waves. The relation (factor 2) between their energies along [100] and [111] directions leads to the conclusion that the coupling along {\bf c} involved in these modes is nul, as shown below in the discussion. In this large q-range, the 2D character persists down to 11K (compare the three upper branches along [111] of (Fig. 4) with those  along [100] reported in Ref 6).
\begin{figure}[t]
\centerline{\includegraphics[width=9 cm]{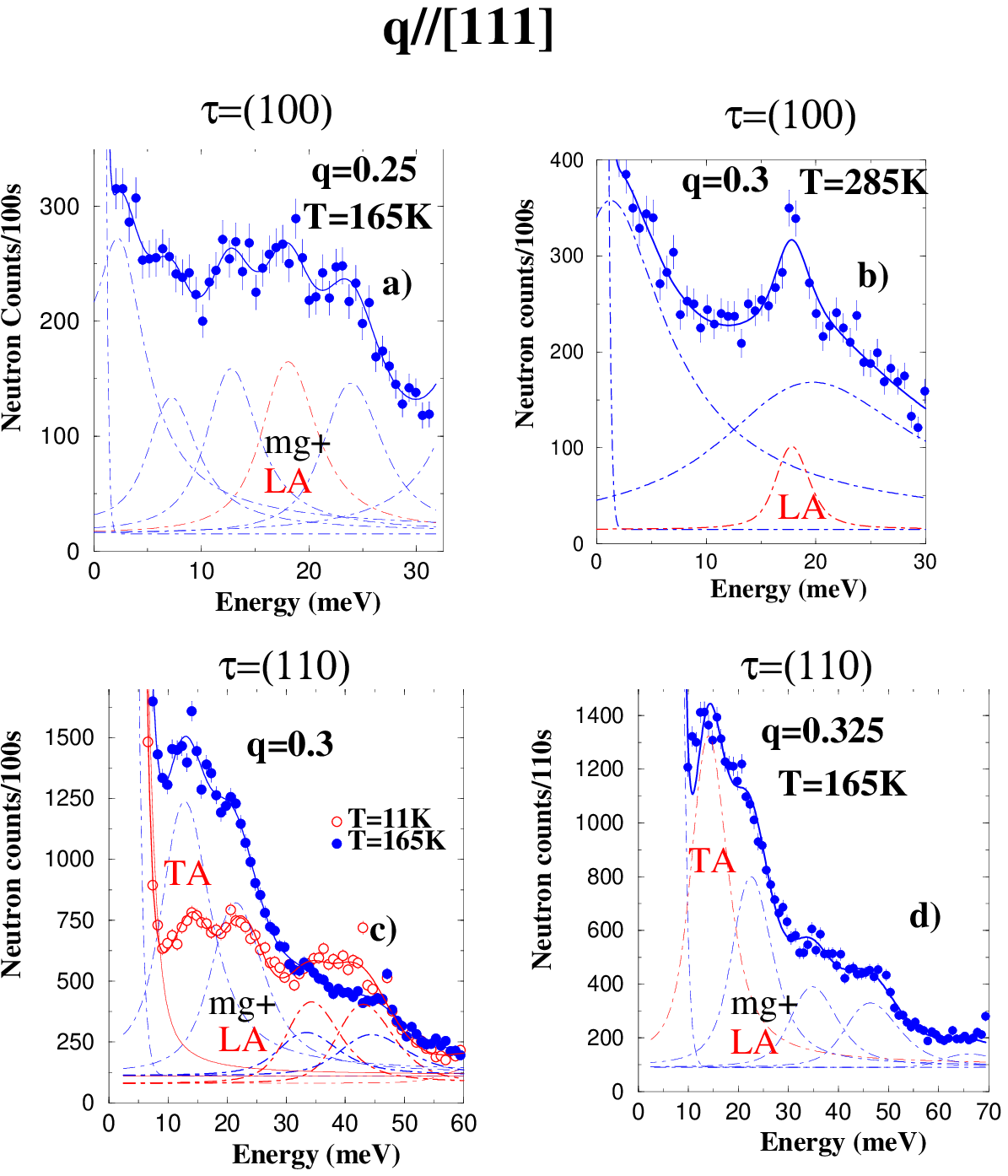}}
\caption{Energy spectra along [111] corresponding to two tranfer energy ranges [0-30meV] and [0-60meV], for which distinct experimental conditions are used: k$_f$=2.662$\AA^{-1}$ around $\tau$=(001) for [0-30meV] and k$_f$=4.1$\AA^{-1}$ around $\tau$=(110) for [0-60meV]. 
In the [0-30meV] range, comparison of the spectra in {\bf a)} at q=0.25, T=165K, where 4 peaks are identified, and in {\bf b)} at the close value q=0.3, T=285K, where only one small peak persists, insures the magnetic nature of the modes of {\bf a)}, with a contamination of the mode at 18 meV by the LA phonon. In the [0-60meV] range, comparison of the two spectra in {\bf c)} at q=0.3, showing an {\it increase of intensity} of the 2 upper modes from T=165K (filled circles) down to T=11K (empty circles), ascertains their magnetic nature. In {\bf d)}, example of spectrum at q=0.325,T=165K.}
\end{figure}
\begin{figure}[t]
\centerline{\includegraphics[width=9 cm]{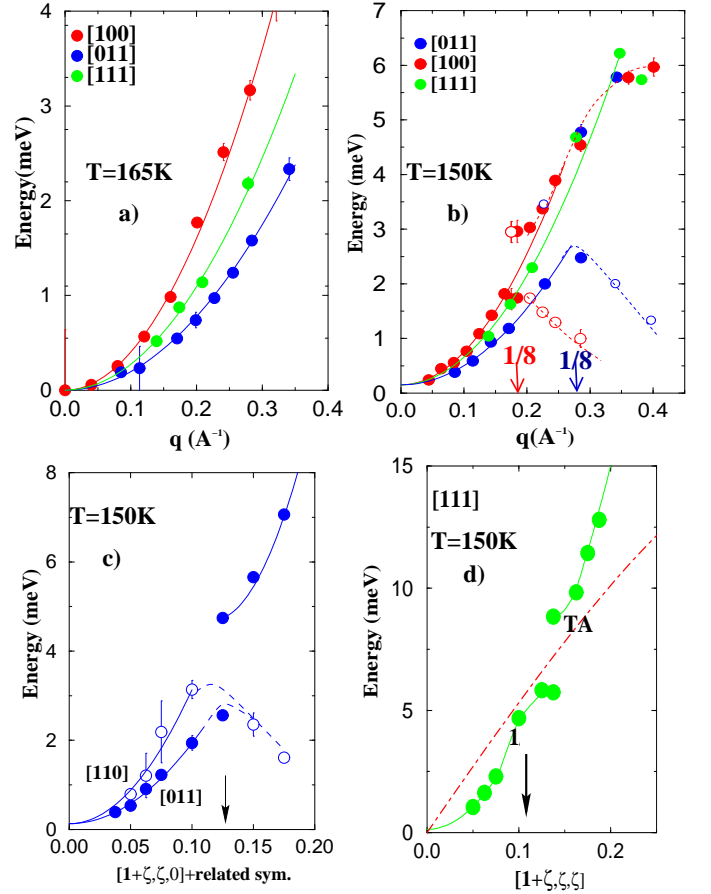}}
\caption{Comparison of the dispersed spin waves along several directions. 
Red, blue and green circles correspond to [100], [011] and [111] directions respectively with weak (empty circles) and main (filled circles) intensities. {\bf a)} at T=165K and {\bf b)} at T=150K in q($\AA^{-1}$). In both panels, the solid lines are fit with $\omega$=Dq$^2$+$\Delta$ laws. In {\bf b)}, arrows indicate gaps at q$_0$$\approx$1/8 rlu (or q$^{'}_0$=0.19$\AA^{-1}$,$\sqrt 2$q$^{'}_0$ and $\sqrt 3$q$^{'}_0$ along [100], [110]+[011] and [111] resp.), at corresponding energies 2.5, 3.6 and 7 meV for the center of the gap. 
{\bf c)} and {\bf d )}: spin wave branches in q(rlu) units, T=150K. {\bf c)}: along [110] and [011]+[101], distinct for q$<$1/8. {\bf d )}: along [111]. The TA phonon branch measured atT=250K is also reported.}
\end{figure} 

The 2D character of the magnetic coupling is also found when considering the small-q range, where the ambiguities related to twinning can be solved. In Fig. 6-a) the quadratic dispersed curves, E(q)=Dq$^2$+$\Delta$, measured in various directions with a good resolution (k$_f$ varying from 1.3 to 1.5$\AA^{-1}$), are compared. Besides D$^{[111]}$=25$\pm$2 meV$\AA^{2}$ for which the q-direction is unambiguous (all domains are equivalent), the largest value, 40$\pm$2 meV$\AA^{2}$ is readily attributed to D$^{[100]}$=D$^{[010]}$ and the lowest one, 19$\pm$2meV$\AA^{2}$ to D$^{[011]}$ =D$^{[101]}$ with $\Delta\approx$0.
  These values obey interpolation relations between the two extreme D$^{[100]}$ and D$^{[001]}$ values, namely D$^{[111]}$=1/3(2D$^{[100]}$+D$^{[001]})$ and D$^{[101]}$=D$^{[011]}$=1/2(D$^{[100]}$+D$^{[001]}$). This leads to the conclusion that D$^{[001]}$$\le$4 meV$\AA^{2}$ along {\bf c}. 
In agreement with this conclusion, the weak magnetic modes along [110], not reported in Fig. 6-a) for clarity, leads to D$^{[110]}$=2D$^{[011]}$.\\

{\bf Evolution of spin waves through the transition (T$_{O'O"}$=159K)}

\begin{figure}[t]
\centerline{\includegraphics[width=8 cm]{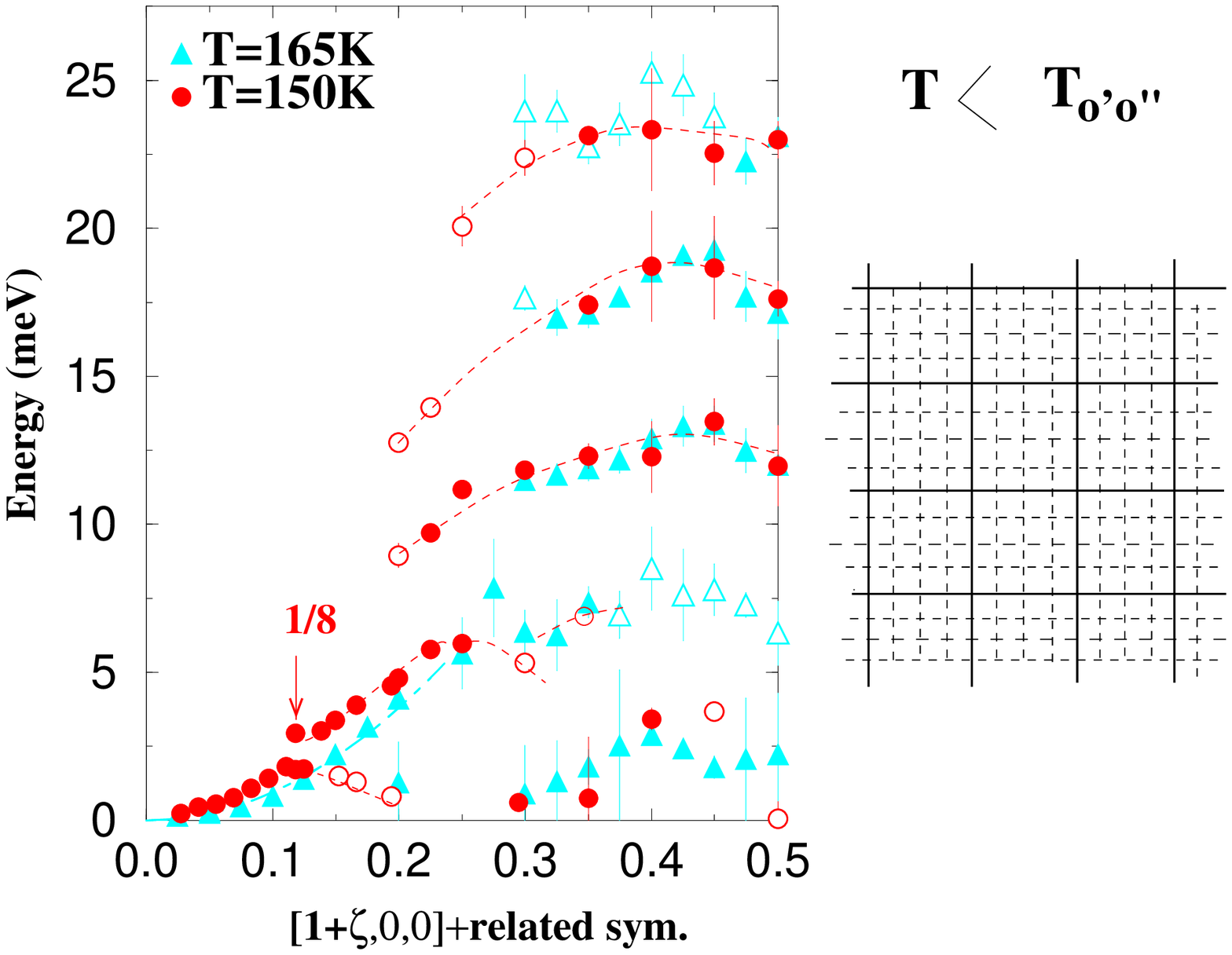}}

\caption{Left panel: spin wave dispersion along [100]+[010]+[001] at T=150K (red circles) and T=165K (cyan circles). Full (empty) circles indicate main (weak) intensity. Dashed lines are guides for the eye. Right panel: example of 2D stripe supertructure with 4a x 4a domains, coupled along {\bf a} and {\bf b}. The hole poor Mn sites are at the intersection of the dashed lines. Bold lines visualize hole-rich lines.  
}
\end{figure}
{\it In the low-q range}, when cooling down through T$_{O'O"}$, 
the most salient feature is the opening of a gap at {\bf q$_0$$\approx$0.125} observed along all symmetry directions (see Fig. 2-a at q=0.127). A folding
of the spin wave branch is specially well observed along [100]+[[010] and [011]+[101] (Fig. 6-b) and Fig. 6-c) respectively). At T=150K, the D values have increased, keeping a quasi 2D character as long as q$<$1/8, (D=66$\pm$2, 36$\pm$2 and 48$\pm$2 meV$\AA^2$ along [100], [011] and [111] respectively, and  $\Delta$=0.13 meV as shown in Fig. 6-b) and 6-c)). In addition, along [100], the damping of the modes has decreased by a factor 1/3 (Fig 2-a). The behaviour of the spin waves propagating along {\bf c} is puzzling. From the above D values, one deduces D$^{001}\approx$5 meV $\AA^2$ for q$<$1/8. However, along [111], the center of the gap, E({\bf q$_0$}=0.125,0.125,0.125) is close to 3 times that, E({\bf q$_0$}=0.125,0,0) along [100]+[010]+[001], indicating a nearly isotropic coupling (Fig. 6-d). Moreover, around the q$_0$=(0.125,0,0) value, some scans indicate an additional mode with a small intensity not reported here, that could be assigned to [001]. This surprising q-evolution along {\bf c} suggests a
magnon-phonon coupling around q$_0$. Interestingly, along [111], at T$_{O'O"}$, the TA phonon branch precisely crosses the magnon branch at these q and energy gap values (Fig. 6-c)). This suggestion will be checked later. Below 150K, in this small q-range, the whole dispersed branch evolves towards isotropy, as observed at 14K\cite{Moussa}. Clearly, the magnetic coupling revealed by the excitations measured in this small-q range and by those of the large q-range cannot be the same, possibly associated with mainly double-exchange and mainly super-exchange respectively.\\
 The opening of the gap and the folding reveal a new periodicity of 4 lattice units along all symmetry directions. This new periodicity is discussed in terms of a new inter-domain coupling in the discussion below.\\
 Although the transition is mainly revealed at q$\approx$1/8, changes also concern {\it the large q-range}. The whole magnetic spectrum obtained at 150K  is compared to that at 165K along [100] in Fig 7. At q=0.3 along [100], the lower mode, less intense, slightly decreases and looks connected to the dispersed branch which shows now a down turn at q=0.25 (cf raw data in Fig. 2-d and Fig. 7) or a new folding. This is consistent with the $\approx$4a periodicity indicated by the folding at q$\approx$0.125. More generally, there is a transfer of spectral weight from the quasi-elastic peak and the lowest energy  level to the two upper levels(Fig. 2-f), 2-h), so that the lowest energy modes have disappeared at T=11K. In addition, the upper levels are now extended towards a q value smaller than q=0.25,  with a beginning of a F dispersion. For example, at q=0.25, Fig. 2-b), the single mode observed above T$_{O'O"}$ becomes divided into several modes below. As observed at 165K, the strong evolution induced by an applied field, which mainly affects the spectral weight of the two modes, corresponds to that observed by lowering temperature (cf Fig. 3, right panel).

{\bf Discussion}

In the metallic state, the q-independent levels do not coincide with the dispersed phonon branches (dot-dashed lines in Fig. 1) except at q=0.5 where the two upper levels lie on LO and LA energy values. Their origin cannot be a magnon-phonon coupling even though this coupling exists\cite{Moussa}. They reveal local magnetic excitations. 
 We propose an interpretation in terms of confined spin waves as in La$_{1-x}$Ca$_x$MnO$_3$\cite{Hennion}, characteristic of domains with a limited size. Compared to La$_{0.83}$Ca$_{0.17}$MnO$_3$, which exhibits a badly defined metal-insulator transition, the new features observed here, such as several magnetic levels in all directions and the well-characterized gap at q$\approx$1/8 rlu, allow a deeper insight into the transition. 
   
 The expression of spin-wave energy in a uniform F state using a 3D Heisenberg model writes: 
 E(q$_a$,q$_b$,q$_c$)=8S\{J$_{ab}$[1-0.5(cos(2$\pi$q$_a$)+cos(2$\pi$q$_b$))]+
J$_{c}$/2[1-cos(2$\pi$q$_c$)]\}. Here, J$_a$=J$_b$=J$_{ab}$ and J$_c$ are the first neighbor coupling along {\bf a} (or {\bf b}) and {\bf c} respectively. The same expression may be used for independent clusters\cite{Hendriksen}.
Along [111], the observation of a factor two between the energy levels with  those along [100], determines a magnetic coupling J$_c$$\approx$0. This results from the compensation of the super-exchange coupling, AF along {\bf c} in pure $LaMnO_3$ by the F coupling induced by the mobile charges, in agreement with predictions\cite{Feiner}. 
Therefore, we conclude that these domains have a 2D character.
Moreover, we privilege a model where the size is limited along {\bf a} and {\bf b} in a same plane rather than along one direction only, for which one would have observed a dispersed spin wave branch at large q, in addition to the dispersionless levels. 
In a limited size cluster of spins, one obtain as many energy levels as spins, but, due to symmetries, the intensity of some of them may vanish along specific cristallographic directions.\\
{\it In the metallic state} where the dispersed branch is defined up to q=0.25,
we suggest to relate the observed energy levels to 4a x 4a limited size domains. Actually, a preliminary calculation of magnetic excitations in such square clusters with boundaries along {\bf a} and {\bf b} performed at T=0, indicate three dispersionless levels along [100] and [010]. The introduction of some disorder, as well as the underlying coupling with phonons may explain some differences with the present observations. A detailed comparison with experiment will be reported elsewhere\cite{Petit}.
  The existence of domains is attributed to a charge segregation\cite{Moreo}. The coupling involved in the large q-range, strongly anisotropic down to T=11K, is typical of the parent compound LaMnO$_3$ which suggests to
characterize the domains as hole-poor, insulating. Therefore, one expect that their boundaries are hole-rich. 
This description implies a very weak inter-cluster coupling across the hole-rich boundaries.   
The existence of a dispersed branch in the small q range, characterized by a D stiffness constant indicates that the picture of independent domains is not correct. It reveals a long-distance magnetic coupling through the domains. Its 2D character suggests that the conductivity has also a 2D character. This dispersed branch, which appears with the long-range magnetic order at T$_C$, is also the most sensitive to the T$_{O'O"}$ transition.  \\

{\it At the T$_{O'O"}$ transition}, the small-q dispersed branch is folded at q$\approx$0.125, with a strong decrease of the spin wave damping along [100]. This reveals a new $\approx$4a periodicity along {\bf a} and {\bf b}. This is a  new result. Up to now, only the direction {\bf c} has been considered in the models of charge ordering. The absence of any reported static superstructure at q=0.25 within the ({\bf a, b}) plane, of atomic or magnetic origin suggests a small contrast of charges inside the clusters and on the boundaries, which is a general feature in manganites.  
To reconcile the picture of charge segregation with 4a limited-size domains in the quasi-metallic state and the new periodicity of 4a which occurs at the transition, we suggest that the domains, separated by fluctuating hole-rich paths above T$_{O'O"}$, become coupled and ordered below T$_{O'O"}$. In the present case where the domains have an equal size (4a) along {\bf a} and {\bf b} in the same plane, this suggests a 2D structure of orthogonal stripes as shown in Fig 7. They should be stabilized by some magnon-phonon coupling\cite{Moussa}, in agreement with predictions\cite{Khomskii}. A preliminary calculation of 2D spin waves in a model where two magnetic couplings are introduced, generating lines with a 4a periodic coupling along {\bf a} and {\bf b}, supports this picture\cite{Petit}. This would be the first observation of stripes in a ferromagnetic state. A similar superstructure with diagonal stripes has been proposed for cuprates\cite{Fine} as an alternative model to the 2D supertructure of parallel stripes\cite{Tranquada}, for which charge localisation is expected\cite{Hasselmann}. There, the domains are antiferromagnetic with an antiphase at the boundaries. The spin waves, calculated for parallel stripes only\cite{Kruger}, can be interestingly compared to the present observations. At low energy, the spin wave branch disperses starting from the magnetic superstructure. Here, it corresponds to the dispersed branch which starts from the underlying q=0.25 superstructure as well as from q=0, as expected in the ferromagnetic case. In the AF case, depending on the strength of the coupling between the AF domains, the spectrum with main intensity consists of one branch only (strong coupling), as found in La$_{1.69}$Sr$_{0.31}$NiO$_4$\cite{Bourges}, or, one branch and q-modulated levels at higher energy (weak coupling). This latter case is close to the present situation. \\   
In conclusion, this spin dynamics study reveals that the quasi-metallic state is well-described by a charge segregation picture, with 2D domains of limited-size in the ({\bf a, b}) plane and is very close to the low-temperature state. This suggests to describe the metal-insulator transition as an ordering of the domain boundaries mainly in the ({\bf a, b}) plane, leading to a 2D stripe supertructure. This would be the first observation of a stripe supertructure in a ferromagnetic case.  

The authors are very indebted to Louis-Pierre Regnault and A. Ivanov for their help in the experiments with polarised neutrons. They also aknowledge S. Petit, G. Bouzerar, B. Canals, B. Hennion, Y. Motome and T. Ziman for fruitful discussions.

\end{document}